  \documentclass[12pt]{article}

\usepackage{epsf}
\usepackage[dvips]{graphicx}
\begin{document}
\begin{center}
{\bf\large{$\Lambda$ AND $K^0_s$ PRODUCTION IN pC COLLISIONS AT 10
GeV/c}}

\vskip 2mm P.Zh.Aslanyan$^{1,2 \dag}$, V.N.Emelyanenko$^1$, G.G.
Rikhkvitzkaya$^1$

 vskip 5mm
{\small (1) {\it Joint Institute for Nuclear Research }
\\
(2) {\it Yerevan State University }
\\
$\dag$ {\it E-mail: paslanian@jinr.ru }}

\end{center}

\begin{abstract}
\begin{minipage}{150mm}
 The experimental data from the 2m propane bubble chamber  have
been analyzed for pC$\to \Lambda(K^0_s)X$ reactions at 10 GeV/c.
 The estimation of experimental inclusive cross
sections for $\Lambda$ and $K^0_s$ production in the p$^{12}C$
collision is equal to $\sigma_{\Lambda}$= 13.3$\pm$1.7 mb and
$\sigma_{K^0_s}$= 3.8$\pm$0.6 mb, respectively.

 The measured $\Lambda /\pi^+$ ratio from pC reaction
 is equal to (5.3$\pm0.8)*10^{-2}$. The experimental $\Lambda /\pi^+$
 ratio in the pC reaction is approximately two times larger than the
 $\Lambda /\pi^+$ ratio  from  pp reactions or from  simulated pC
 reactions by FRITIOF model for the same energy. The $\Lambda
/\pi^+$ ratio  in interaction C+C at momentum 10 Gev/c is four
times larger than the $\Lambda/\pi^+$ ratio from p+p reactions at
the same energy.

The investigation has been performed at the Veksler and Baldin
Laboratory of High Energies, JINR.
\end{minipage}
\end{abstract}

\section{Introduction}

  Strangeness enhancement has been extensively discussed as a possible
signature for the quark-gluon plasma(QGP)\cite{1,2}. Strange
particle production has also been analyzed regarding such reaction
mechanisms as the multinucleon effect\cite{3}, or the fireball
effect\cite{4}, or as the deconfiment signal, within the context
of thermal equilibration models\cite{5}-\cite{8}.

In particular, strange particles have been observed extensively on
hadron - nucleus and nucleus-nucleus collisions 4-15 Gev
regions\cite{9}-\cite{14}. The strange hyperon
yields\cite{9}-\cite{11} are therefore of great interest as an
indicator of strange quark production. The number of $\Lambda$s
produced in $\overline{p}$+Ta reaction at 4 GeV/c was 11.3 times
larger than that expected from the geometrical cross section
\cite{9}. Experiments with Si+Au and Au+AU collisions  at
11.6\cite{13} and 14.6 A GeV/c \cite{14} measured a $K^+/\pi^+$
ratio in heavy-ion reactions that is four to five times larger
than the $K^+/\pi^+$ ratio from p+p reactions at the same energy.
The thermal model\cite{6} gives an good description of
$K^+/\pi^+$, $\Lambda /\pi^+$ ratio for data Au+Au, Si+Au
interaction at momenta 10-15 A Gev/c and, showing a broad maximum
at the same energies.

However, there have not been sufficient experimental data
concerning strange-hyperon production over 10-40 GeV/c momentum
range in proton - nucleus and nucleus-nucleus reactions. In this
paper the new results are presented the measured inclusive cross
sections for $\Lambda(K^0_s)$ production and $\Lambda/\pi^+$ ratio
in the reaction p+$^{12}C$.

\section{  Experimental procedure }

\subsection{METHOD}

 Searching for the $V^0$  on $\approx$ 700000(or 345*2 tapes) photographs
 of the JINR 2m propane bubble chamber exposed to a 10GeV/c proton beam
 \cite{16}-\cite{21}.  The primary proton beams must be to
 satisfy of conditions: $\mid tg\alpha\mid <0.02$ 1.62$<\beta <$1.69 rad.
 The magnetic field (B=15.2 kG) measurement
 error is $\Delta$B/B=1\%. The fit GRIND -based program
GEOFIT\cite{18} is used to measure the kinematics track parameter
p, $\alpha$, $\beta$. Measurements were repeated three times for
events which failed in reconstruction by GEOFIT.

The estimate of ionization, the peculiarities of the end track
points of stopping particles permitted one to identify them over
the following momentum ranges : protons of $0.150\le p\le $ 0.900
GeV/c and $K^{\pm} of p\le$0.6 GeV/c.

\subsection{Identification of $\Lambda$ and $K^0_s$ }

 The events with $V^0$ ($\Lambda$ and $K^0_s$)  were identified
using the following criteria \cite{19,20}:\\
1) $V^0$ stars from the photographs were selected according to
$\Lambda\to\pi^-+p$, neutral $K_s\to\pi^-+\pi^+$ or $\gamma \to
e^++e^-$ hypothesis. A momentum limit of $K^0_s$ and $\Lambda$ is
greater than 0.1 and 0.2 GeV/c, respectively ; 2) $V^0$ stars
should have the effective mass of $K^0_s$ and of $\Lambda$; 3)
these $V^0$ stars are directed to  some vertices(complanarity); 4)
they should have one vertex, a three constraint fit for the $M_K$
or $M_{\Lambda}$ hypothesis  and after the fit, $\chi^2_{V^0}$
should be selected  over range less than 12; 5)The analysis has
shown\cite{20} that the events with undivided $\Lambda K^0_s$ were
assumed to be events as $\Lambda$.

Table~\ref{tab:evol} presents (70\%)the number of experimental
$V^0$ events produced from interactions of: a) primary  proton
beams, b)secondary charged particles and c)secondary neutral
particles.

The $V^0$s classified into three grades. The first grade comprised
$V^0$s  which could be identified with above cuts and bubble
densities of the positive track emitted from the $V^0$s. The
second grade comprised $V^0$s which could be undivided $\Lambda
K^0_s$. For correctly identification the undivided $V^0$s are used
the $\alpha$(Fig.1a) and the $cos\theta^*_{\pi^-}$(Fig.1b)
distributions.
$$
\alpha =
(P^+_{\parallel}-P^-_{\parallel})/(P^+_{\parallel}-P^-_{\parallel})
$$
Where $P^+_{\parallel}$ and $P^-_{\parallel}$  are the momentum
components of  positive and negative charged tracks from the
$V^0$s  relative direction of the $V^0$s momentum.The

$cos\theta^*_{\pi^-}$ is the angular distribution of $\pi^-$ from
$K^0_s$ decay. The $\alpha$(Fig.1a) and the $cos\theta^*_{\pi^-}$
distributions from $K^0_s$ decay were isotropic in the $K^0_s$
rest frame after removing undivided $\Lambda K^0_s$. Then these
$\Lambda K^0_s$ events appropriated events as $\Lambda$. After
 we show in Fig.1c that the $cos\theta^*_{\pi^-}$
distributions for the $\Lambda$+$\Lambda K^0_s$s have been also
isotropic in $V^0$ rest frame. As a result of above procedure have
lost of $K^0_s$ 8.5\% and admixture of $K^0_s$ in $\Lambda$s
events 4.6\%. The third grade comprised $V^0$s  which could be the
invisible $V^0$s at a large azimuth angle $\phi$\cite{20}. The
average $\phi$ weights   were $<w_{\phi}>$ = 1.06$\pm$0.02 for
 $K^0_s$ and $<w_{\phi}>$ = 1.14$\pm$ 0.02 for $\Lambda$.

Figures.2a,c and 2b,d show the effective mass distribution of
$\Lambda$(8657-events),$K^0$(4122-events) particles and  their
$\chi^2$ from kinematics fits, respectively, produced from the
beam protons interacting with propane targets. The measured masses
of these events have the following Gaussian distribution
parameters $<M(K_s)>$= 497.7$\pm $3.6, s.d.= 23.9 MeV/$c^2$ and
$<M(\Lambda)>$ =1117.0$\pm$ 0.6, s.d.=10.0 MeV/$c^2$. The masses
of the observed $\Lambda$, $K^0_s$ are consistent with their PDG
values. The expected functional form for $\chi^2$  is depicted
with the dotted histogram(Fig.2).

 Each $V^0$ event weighted by a factor $w_{geom}$ (=1/$e_{\tau}$), where $e_{\tau}$
is the probability for potentially observing the $V^0$, it can be
expressed as
$$
e_{\tau}= exp(-L_{min}/L)- exp(-L_{max}/L),
$$
where L(=cp$\tau$/M) is the flight length of the $V^0$,$L_{max}$
the path length from the reaction point to the boundary of
fiducial volume, and $L_{min}$(0.5 cm) an observable  minimum
distance between the reaction point and the $V^0$ vertex.
M,$\tau$, and p are the mass, lifetime, and momentum of the $V^0$.
The average geometrical weights were 1.34$\pm$0.02 for $\Lambda$
and 1.22$\pm$0.04 for $K^0$.

Now, let us examine a possibility from neutron stars of imitating
 $\Lambda$ and $K^0_s$ the using model FRITIOF\cite{22} for the hypotheses
reaction p+C$\to$n+X,n+n$\to \pi^- p(or \pi^- \pi^+) +X^0$  with
including fermi motion in carbon. Then, these background events
were analyzed by using the same experimental condition for the
selection $V^0$s. The 2 vertex analysis have shown the background
from neutron stars are equal to 0.1\% for $\Lambda$ and 0.001 for
$K^0_s$ events.

\subsection{The selection of interactions on carbon nucleus}

The criteria for selection of interaction with carbon has
shown\cite{19,25}. The  p+C$\to \Lambda(K^0_s)X$ reaction were
selected by the following criteria:

 1. Q = $n_+ ~ - ~n_- >$ 2; \\
 2. $n_p~+~n_{\Lambda}>$1;\\
 3. $n_p^b ~+ ~n_{\Lambda}^b>$0; \\
 4. $n_->$2;\\
 5. $n_{ch}$= odd number ; \\
 6. $\frac{E_p(\Lambda)-P_{p(\Lambda)}cos\Theta _{p(\Lambda)}}{m_t}>$1.

$n_+ ~and ~n_ ->$ are the number of positive and negative
particles on the star;$n_p~ and ~n_{\Lambda}$ are the number
protons and $\Lambda$ hyperons with momentum p$<$0.75 GeV/c on the
star.$n_p^b$ è $n^p_{\Lambda} $ are the number protons and
$\Lambda$ hyperons  to emitted in backward direction.
$E_{p(\Lambda)},~P_{p(\Lambda)}$ and $\Theta _{p(\Lambda)}$ are a
energy, a momentum and a emitted angle of protons(or $\Lambda$s)
in the Lab. system. $m_t$ is the mass of target. These criteria
were separated  $\approx$ 83 \% from all inelastic p+C
interactions\cite{25}.  The p+C events were selected by the above
criteria the using FRITIOF model \cite{22}. Results of the
simulation have lost 18\% and 20\% from interactions pC$\to
\Lambda$ X and pC$\to K^0_s$X, respectively. The contribution from
pp$\to \Lambda$ X and pp$\to K^0_s$X in pC interactions  are equal
to 1.0\% and 0.3\%, respectively.

\section{The measured cross sections $\Lambda$ and $K^0$}

The cross section is defined by the formula:
$$
  \sigma = \frac{\sigma_0}{e}\prod_iw_i=
\frac{\sigma_r*N_r^{V^0}*w_{hyp}*w_{geom}*w{\phi}*w_{kin}*w_{int}}{N_{int}^r*e_1*e_2*e_3},
(3.1)
$$
 where  $e_1$
is the efficiency of search for $V^0$ on the photographs, $e_2$
the efficiency of measurements. The $V^0$s of 75\%(preliminary)
could be successfully reconstructed and accepted in the analysis.
$e_3$ the probability of decay via the  channel of charged
particles ($\Lambda\to p\pi^-, K^0\to\pi^+\pi^-$), $\sigma_0=
\sigma_r/N_r$ the total cross section, where  $\sigma_r $ is the
total cross section for registered events, $N_ r$ is the total
number of registered interactions of beam protons over the range
of the chamber.
 $\sigma_t(p+C_3H_8)=3\sigma_{pC}+8\sigma_{pp}$=(1456$\pm$88)mb \cite{27},
where $\sigma_t,\sigma_{pC}~ and~ \sigma_{pp} $ are the total
cross sections in interactions $p+C_3H_8$,p+C and p+p,
respectively. The propane bubble chamber method have been
permitted the registration the part of all elastic interactions
with the propane \cite{23,24}\, therefore the total cross section
of registered events is equal to:
$\sigma_r(p+C_3H_8)=3\sigma_{pC}(inelastic)+
8\sigma_{pp}(inelastic)+8\sigma_{pp}(elastic)0.70=(1049\pm60)$mb.

$w_i$ are  weights for the lost events with $V^0$
for(Table~\ref{tab:weight}): $w_{geom}$ - the $V^0$ decay outside
the chamber; $w_{\phi}$ - the required isotropy for $V^0$ in the
azimuthal (XZ) plane; $w_{hyp}$ - the undivided $\Lambda K^0_s$
events; $w_{int}$ - the selected as $p~+~^{12}C$ from the
interaction of $p~+~C_3H_8$; $w_{kin}$ - the kinematic
conditions(with FRITIOF);$w_{int}$- the $V^0$+ propane
interactions.

Table~\ref{tab:evcros} show that the experimental cross sections
are calculated by formula 3.1  for inclusive $\Lambda$ hyperons
and $K_s^0$ mesons  productions in the interactions of pp and pC
at beam momentum 10 GeV/c.

 Ratios of average  multiplicities
$\Lambda$ hyperons and $K_s^0$ mesons to  multiplicities $\pi^+$
mesons  in p+C interaction at beam momenta  4.2 GeV/c and 10 GeV/c
show in Table~\ref{tab:multpc}.  Experimental data on
multiplicities $\pi^+$ mesons in the interactions of pC at momenta
4.2 GeV/c ($<n_{\pi+}>=0.71\pm $0.01) and 10 Gev/c
($<n_{\pi+}>=1.0\pm $0.05) taken from publications \cite{26} and
\cite{25}, respectively.

The $\Lambda /\pi^+$ ratio for C+C reaction is shown in
Table~\ref{tab:multcc} and on Fig.3. This ratio have been obtained
by using the Glauber approach  on the experimental cross section
for p+C $\to\Lambda $X reaction.

           \section{Conclusion}

The experimental data from the 2 m propane bubble chamber  have
been analyzed for pC$\to\Lambda(K^0_s)$X reactions at 10 GeV/c.
The estimation of experimental inclusive cross sections for
$\Lambda$ and $K^0_s$ production in  pC collisions is equal to
$\sigma_{\Lambda}$= 13.3$\pm$1.7 mb and $\sigma_{K^0_s}$=
3.8$\pm$0.6 mb, respectively. The measured $\Lambda /\pi^+$ ratio
in pC and pp reactions is equal to  (5.3 ± 0.8)*10-2 and (2.7 ±
0.4)*10-2, respectively. The experimental $\Lambda /\pi^+$ ratio
in the pC reaction is  approximately two times larger than the
$\Lambda /\pi^+$ ratio  from  pp reactions or from simulated pC
reactions by FRITIOF model  for the same energy.The $\Lambda
/\pi^+$ ratio in C+C collisions at 10.0 A GeV/c obtained that is
four times larger than the $\Lambda /\pi^+$ ratio from p+p
reactions at the same energy.

\begin{table}
\caption{The amount (70 \%) of $V^0$ events from interactions of different types which were registrated on photographs with propane bubble chambers method.}
\label{tab:evol}
\begin{tabular}{|l|l|l|l|l|l|}  \hline
 & \multicolumn{3}{|l|}{The amount events from interactions.:}& Total  \\
Chanel&primary beam&sec. charged&sec. neutral&events\\
      &protons &particles & particles & \\ \hline
$\to \Lambda(only) x$&5276&2814&1063&9387\\ \hline
$\to K_s^0(only)x$&4122&1795&481&6543\\ \hline $\to (\Lambda$ and
$K_s^0$) x&3381&1095&376&4608\\ \hline
\end{tabular}
\end{table}

\begin{table} \caption{ Weight  of  the  lost experimental events with $\Lambda$ and $K_s^0$ for pC and pp interactions. }
\vskip 1mm
\label{tab:weight}
\begin{tabular}{|l|l|l|l|l|l|l|l|l|l|l|}  \hline
Type of&1/$e_1$&1/$e_2$&$w_{geom}$&$w_{\phi}$&$w_{int}$&$w_{kin}$&$1/e_3$&W$_{sum}$\\
reaction&&&&&&&&\\ \hline pC$\to \Lambda
X$&1.14&1.25&1.34&1.14&1.11&1.18&1.56&4.37$\pm$0.37\\ \hline
pp$\to \Lambda
X$&1.14&1.25&1.36&1.14&1.11&1.37&1.56&5.15$\pm$0.44\\ \hline
pC$\to K^0_s X$
&1.14&1.25&1.22&1.06&1.04&1.04&1.47&2.93$\pm$0.25\\ \hline
pp$\to K^0_s X$ & 1.14&1.25&1.36&1.06&1.05&1.06&1.47&3.31$\pm$0.28\\
\hline
\end{tabular}
\end{table}
\begin{table}
\caption{Cross sections  $\Lambda$ hyperons and $K_s^0$ mesons for
pp and pC interactions at  beam momentum 10 GeV/c. } \vskip 1mm
\label{tab:evcros}
\begin{tabular}{|l|l|l|l|l|l|l|l|l|l|}  \hline
Type of &$N_{V^0}^{exp.}$&$W_{sum}$&$N_{V^0}^t$&$n_{V^0}=N_{V^0}^t/N_{in}$&$\sigma$ \\
reaction& &&Total&&mb\\ \hline pC$\to
\Lambda X$&6126&4.37$\pm$0.37&26770&0.053$\pm$0.005&13.3$\pm1.6$\\
\hline pp$\to \Lambda
X$&836&5.15$\pm$0.44&4303&0.026$\pm$0.003&0.80$\pm$0.08
\\\hline pC$\to K^0_s
X$&3188&2.93$\pm$0.25&9341&0.018$\pm$0.002&3.8$\pm$0.5
\\\hline
 pp$\to K^0_s X$&699&3.31$\pm$0.28&2313&0.015$\pm$0.001&0.43$\pm$0.04\\\hline
\end{tabular}
\end{table}

\begin{table}
\caption{Ratios of average  multiplicities  $\Lambda$ hyperons and $K_s^0$ mesons to  multiplicities $\pi^+$ mesons  for p+C interaction at beam momenta  4.2 GeV/c and 10 GeV/c. }
 \vskip 1mm
 \label{tab:multpc}
\begin{tabular}{|l|l|l|l|l|l|}  \hline
 &pC This &pC &Cp&Cp  \\
&experiment&FRITIOF&Experiment& FRITIOF \\  &(10 GeV/c)&(10
GeV/c)&(4.2 GeV/c)&(4.2 GeV/c)\\  \hline
$<n_{\Lambda}>/<n_{\pi+}>\times10^2$&5.3$\pm$0.8&2.6&0.7$\pm$0.3&0.9\\\hline
$<n_{K^0_s}>/<n_{\pi+}>\times10^2$&1.8$\pm$0.3&1.8&0.3$\pm$0.2&0.3\\
\hline
\end{tabular}
\end{table}

\begin{table}
\caption{Ratios of  average  multiplicities $\Lambda$ hyperons to
multiplicities $\pi^+$ mesons for C+C interactions  at beam
momentum 4.2 and 10 GeV/c.}
 \vskip 1mm
 \label{tab:multcc}
\begin{tabular}{|l|l|l|l|l|l|}  \hline
 &4.2&10 \\
&Experiment&Experiment \\ \hline
$<n_{\Lambda}>/<n_{\pi+}>\times10^2$&2.0$\pm$0.6&10.9$\pm$1.7\\
\hline
\end{tabular}
\end{table}

\newpage
\begin{figure}[ht]
 \centerline{
\includegraphics[width=150mm,height=180mm]{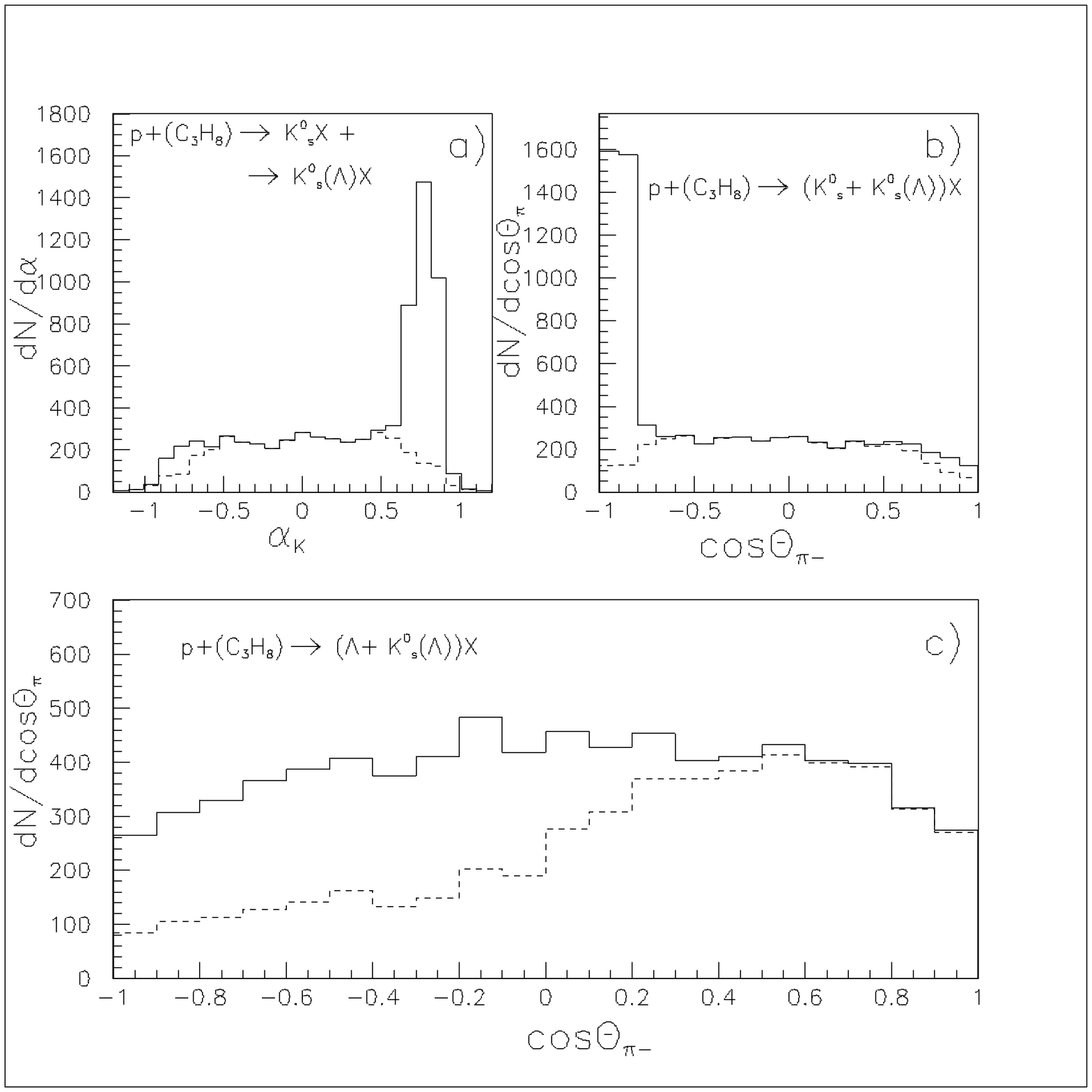}}
 \caption{ Distributions  of  $\alpha$ (Armenteros parameter) and  cos$\Theta^*$-
  are used  for  correctly identification  of   the undivided
  V0s. $\alpha = (P^+_{\parallel}-P^-_{\parallel})/((P^+_{\parallel}-P^-_{\parallel})$. Where $P^+_{\parallel}$
  and  $P^-_{\parallel}$ are the parallel components of momenta  positive and  negative
  charged  tracks. cos$\Theta^*$ - is the angular distribution  of $\pi^-$ from
  $K_s^0$  decay. Distributions of $\alpha$  and  cos$\theta$- were isotropic in the rest  frame
  of $K_s^0$  when  undivided $\Lambda K_s^0$ were assumed to be  events  as $\Lambda$. }

\end{figure}

\begin{figure}[ht]
 \epsfysize=180mm
 \centerline{
 \epsfbox{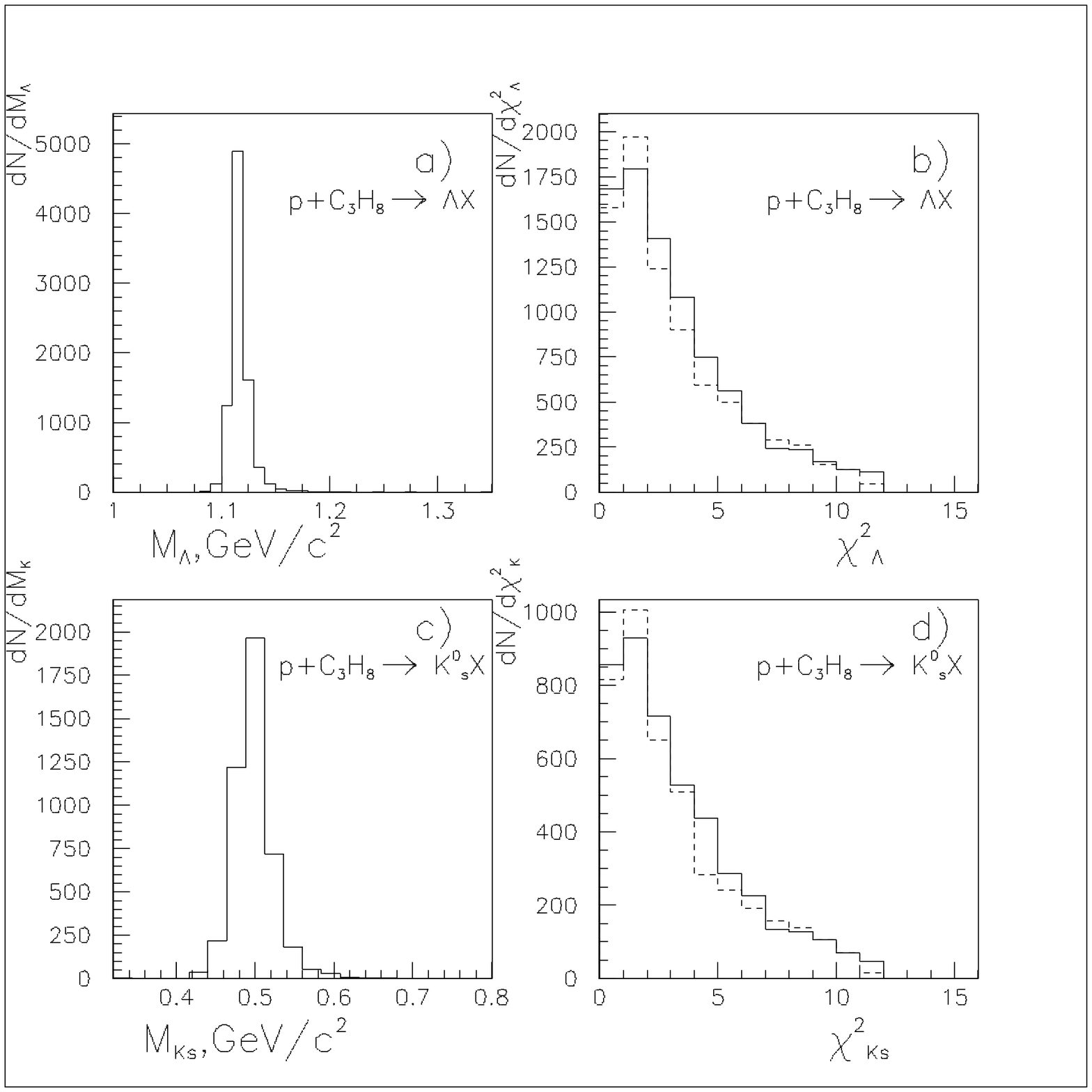}}
 \caption{ The distribution of experimental $V^0$ events produced
 from interactions of beam protons with propane: a) for the effective mass of
  $M_{\Lambda}$; b)for $\chi^2_{\Lambda}(1V-3C)$ of the fits via the decay mode
$\Lambda\to \pi^-+p$; c) for the effective mass of
$M_{K^0_s}$;d)for $\chi^2_{K^0_s}(1V-3C)$ of the fits via decay
mode $K^0_s\to\pi^-+\pi^+$. The expected functional form for
$\chi^2$ is depicted with the dotted histogram.}
\end{figure}

\begin{figure}[ht]
 \centerline{
 \includegraphics[width=150mm,height=160mm]{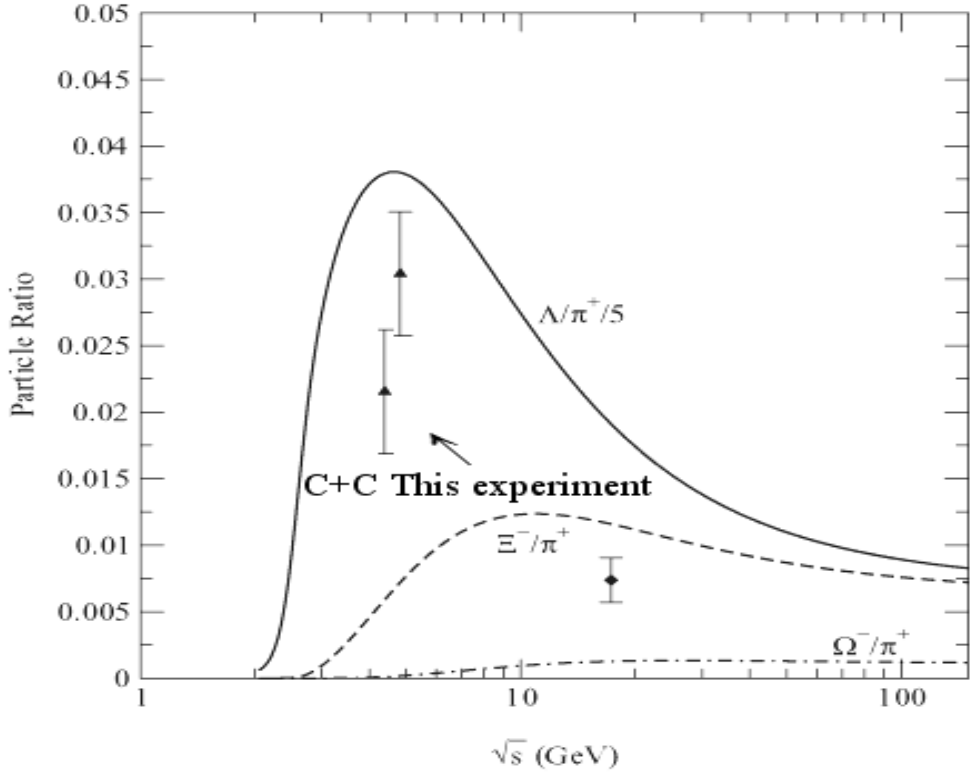}}
 \caption{Prediction of the statistical-thermal model\cite{6} for $\Lambda/\pi^+$(note the factor 5), and
 $\Xi^-/\pi^+$ and $\Omega^-/\pi^+$ ratios a function of
 $\sqrt{s}$. For compilation of AGS data see \cite{7}. The $\Lambda/\pi^+$ ratio in
 interaction C+C on figure is obtained by using data from this experiment. }
\end{figure}

\end{document}